\documentclass[aps,prb,twocolumn,superscriptaddress,showpacs]{revtex4}
\usepackage{graphicx}
\newcommand{\nbsn}{Nb$_{3}$Sn}

\newcommand{\xv}{$\xi_v$}
\newcommand{\msr}{$\mu$SR}

\begin{document}


\title{Magnetic field-induced quasiparticle excitation in Nb$_3$Sn: Evidence
for anisotropic $s$-wave pairing}


\author{R. Kadono$^*$}
\affiliation{Institute of Materials Structure Science, High Energy Accelerator Research Organization (KEK), Tsukuba, Ibaraki 305-0801, Japan}
\affiliation{School of High Energy Accelerator Science, 
The Graduate University for Advanced Studies, Tsukuba, Ibaraki 305-0801, Japan}
\author{K. H. Satoh}
\affiliation{School of High Energy Accelerator Science, 
The Graduate University for Advanced Studies, Tsukuba, Ibaraki 305-0801, Japan}
\author{A. Koda}
\affiliation{Institute of Materials Structure Science, High Energy Accelerator Research Organization (KEK), Tsukuba, Ibaraki 305-0801, Japan}
\affiliation{School of High Energy Accelerator Science, 
The Graduate University for Advanced Studies, Tsukuba, Ibaraki 305-0801, Japan}
\author{T. Nagata}
\affiliation{Department of Physics, Ochanomizu University,
Bunkyo-ku, Tokyo, 112-8610, Japan}
\author{H. Kawano-Furukawa}
\affiliation{Department of Physics, Ochanomizu University,
Bunkyo-ku, Tokyo, 112-8610, Japan}
\author{J. Suzuki}
\affiliation{Advanced Science Research Center, Japan Atomic Energy Agency,
Tokai, Ibaraki 319-1195, Japan}
\author{M. Matsuda}
\affiliation{Advanced Science Research Center, Japan Atomic Energy Agency,
Tokai, Ibaraki 319-1195, Japan}
\author{K. Ohishi}
\affiliation{Advanced Science Research Center, Japan Atomic Energy Agency,
Tokai, Ibaraki 319-1195, Japan}
\author{W. Higemoto}
\affiliation{Advanced Science Research Center, Japan Atomic Energy Agency,
Tokai, Ibaraki 319-1195, Japan}
\author{S. Kuroiwa}
\affiliation{Department of Physics, Aoyama-Gakuin University, 
Sagamihara, Kanagawa, 229-8558 Japan}
\author{H. Takagiwa}
\affiliation{Department of Physics, Aoyama-Gakuin University, 
Sagamihara, Kanagawa, 229-8558 Japan}
\author{J. Akimitsu}
\affiliation{Department of Physics, Aoyama-Gakuin University, 
Sagamihara, Kanagawa, 229-8558 Japan}


\date{\today}

\begin{abstract}

The response of vortex state to the magnetic field in \nbsn\ is probed using
muon spin rotation and small-angle neutron scattering.
A transformation of vortex structure between hexagonal and squared
lattice is observed over a relatively low field range of 2--3 Tesla.  
The gradual increase of the magnetic penetration depth with increasing
field provides microscopic evidence for anisotropic (or multi-gapped) 
$s$-wave pairing suggested by the Raman scattering experiment. 
This result renders need for careful examination on the difference of 
electronic properties between \nbsn\ and V$_3$Si.

\end{abstract}


\maketitle




%
%

While the technologies for utilizing high-$T_c$ cuprates are making steady progress
in recent years,  triniobium stannide (\nbsn) 
is still one of the most important materials for  
the practical application of superconductivity after fifty years since its 
discovery.\cite{Matthias:54} 
Because of its convenient characters such as modestly high transition temperature 
($T_c\simeq18.3$ K) and remarkably high upper critical field ($\mu_0H_{c2}\simeq24.5$ T),
\nbsn\ is used at the heart of superconducting devices, most notably 
as superconducting wires in high-field magnets.
\nbsn\ belongs to a class of $A_3B$ 
binary intermetallic compounds with the A15 or $\beta$-tungsten  
structure, where those based on Nb give rise to a subclass that includes 
Nb$_3$Ge\cite{Matthias:65}, which used to 
dominate the highest $T_c$  ($\simeq$ 23 K) for more than thirty years, 
and therefore it was once a subject of intensive 
study until '70s.\cite{Testardi:75,Weger:73,Allen:80}
Surprisingly, however, it seems that the microscopic details on the superconducting
characters of \nbsn\ are not fully understood to date.  
For example, the revelation of a structural transformation from
cubic to tetragonal with decreasing temperature not far above $T_c$
(martensitic phase transition, which occurs at $T_M=43$ K in \nbsn\cite{Mailfert:67})
led to the argument that the coupling to the instability of lattice vibrations
(soft phonons) associated with the martensitic 
transition might contribute to the increase of $T_c$.
Experimental verification of the elastic softening\cite{Shirane:71,Axe:73}
prompted various models for microscopic mechanism of the martensitic transition, 
many of which emphasize the importance of 
the quasi-one-dimensional character of Nb atoms in the A15 structure
that might also lead to the
anomaly in the electronic density of states,
the Kohn anomaly due to electron-phonon interaction,
or to the Peierls instability.\cite{Testardi:75,Weger:73,Allen:80}
Unfortunately, while these models were partially successful to explain the elastic properties 
of \nbsn, none of them has appeared as unambiguous explanation for the high $T_c$.

\begin{figure}[h]
\includegraphics[width=1.0\linewidth]{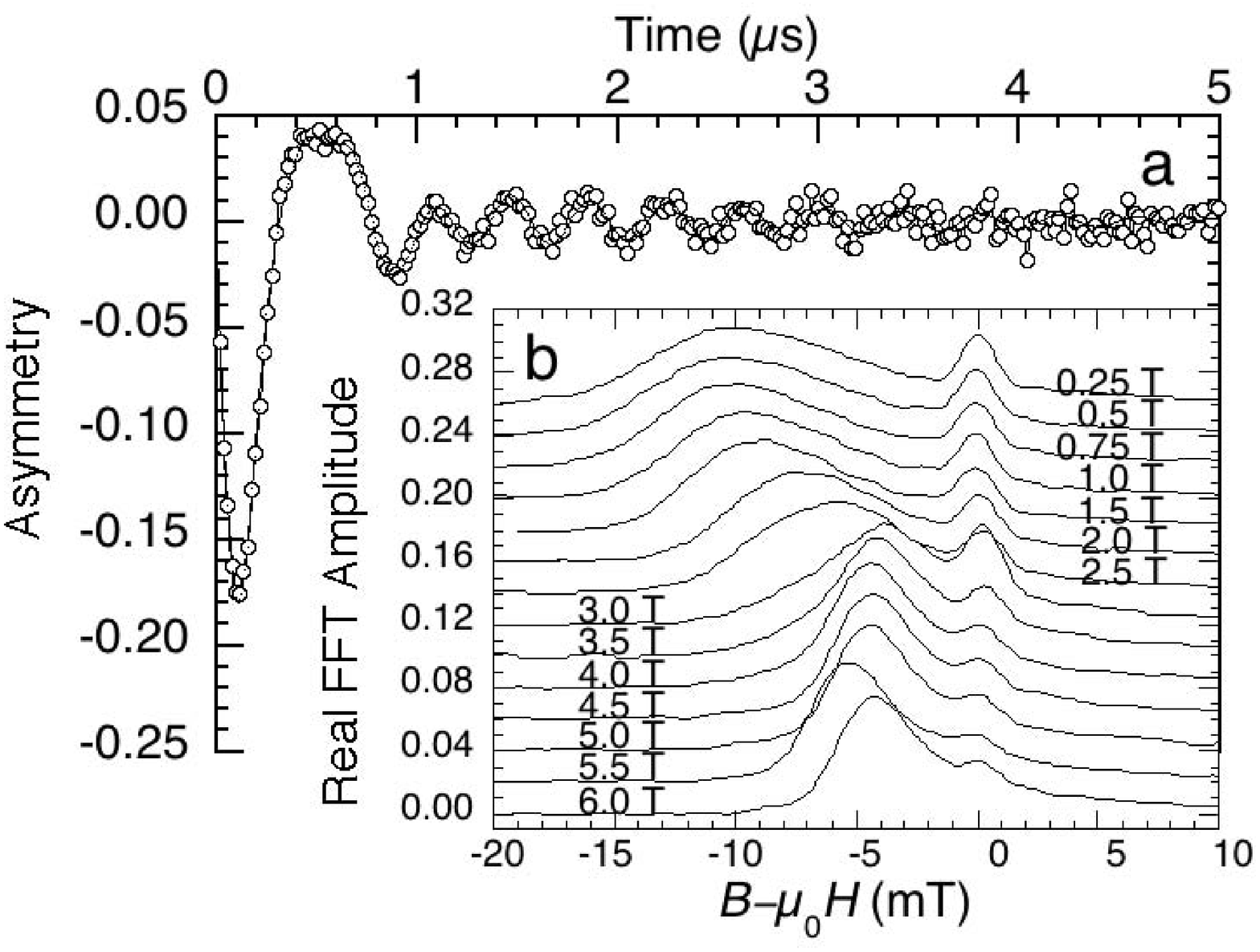}%
\caption{a) Example of \msr\ time spectrum ($\mu_0H=2$ T) 
displayed in the rotating reference frame of 201 MHz (where
solid curve represents fitting result using a model described in the text,
yielding a reduced chi-square of 1.21), and 
b) real amplitude of the fast Fourier transform (FFT, raw data) 
of the time spectra under several magnetic fields in the mixed state of \nbsn\  
($\propto n(B)$), where each spectrum is shifted vertically for clarity.
The background signal from muons stopping in a lucite scintillator (used both
as a sample holder and a muon veto counter) gives rise to the peak at 
$B-\mu_0H=0$, where the amplitude is determined by the
efficiency of the muon veto counter ($\simeq95$ \%, which leads to a relative 
background yield of $\sim5$ \%). 
\label{fftall}}
\end{figure}

Here, we report on the first systematic study of vortex (flux line) state in \nbsn\
using muon spin rotation (\msr) and small-angle neutron scattering (SANS).
The combined use of $\mu$SR and SANS
on the identical specimen is proven to be ideal for the study of the mixed state, 
as they provide complementary information on the magnetic field distribution
associated with flux line lattice (FLL):
\msr\ is sensitive to local field distribution while SANS is sensitive to the long range
structure of FLL.   Moreover, it eliminates much of the problems associated with the 
difference in the preparation of specimen.  
In this brief communication, we show that one of the keys to understand the 
superconductivity of \nbsn\ is the anisotropy in the electronic state, which is 
demonstrated by the transformation of FLL structure from hexagonal to squared
with increasing flux density.  Similar observation in V$_3$Si strongly suggests
that this is a common feature of the superconductors with A15 
structure,\cite{Yethiraj:99,Sosolik:03} where the anisotropy of the 
Fermi surface is reflected through the nonlocal effect.\cite{Kogan:97a,Kogan:97b}
Meanwhile, the field-induced enhancement of the magnetic
penetration depth at lower fields, which is absent in V$_3$Si,\cite{Sonier:04b} 
indicates the presence of low energy quasiparticle
excitation due to the quasiclassical Doppler shift.\cite{Volovik:93} 
This result strongly suggests that the superconducting order parameter also 
has an anisotropic (or multi-gapped) structure on the Fermi surface.\cite{Kadono:04}

\nbsn\ samples were prepared by the method described elsewhere.\cite{Fujii:82}
The superconducting transition temperature was determined to be 
17.9(2) K, as inferred from the shielding effect in magnetization measurement.
Several single crystals with a sizable dimension ($\sim10^2$ mg each, covering
a beam spot area of $7\times7$ mm$^2$ with random crystal orientation) were used 
for \msr\ measurement with a high transverse field spectrometer (HiTime)
installed on the M15 beamline of TRIUMF, Canada, where they were irradiated by 
a muon beam of 4 MeV. 
Further details on the \msr\
measurements using HiTime might be found in earlier publications.\cite{Ohishi:03} 
One of these crystals, 
which was confirmed to be of single domain at ambient temperature, was used  
for the neutron measurements with the SANS-J spectrometer of JRR-3, 
Japan Atomic Energy Agency.  
In both measurements, precautions were
taken to minimize the effect of flux pinning: every time when external magnetic field
was changed, the sample temperature was raised above $T_c$ and cooled down again
after the magnetic field was settled.
The magnetic field dependence was obtained for transverse field \msr\ spectrum 
(TF-\msr, where the initial muon spin polarization is perpendicular to the
 external field) at 2.0--2.5 K and for SANS spectrum at 
 2.6(1) K, respectively.

\begin{figure}[h]
\includegraphics[width=1.0\linewidth]{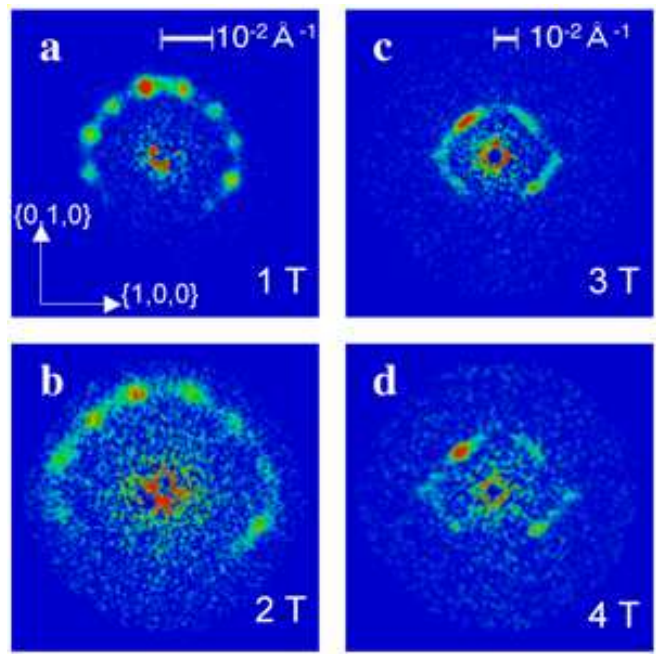}%
\caption{SANS diffraction patterns obtained by subtracting the background data at
30 K from the data taken at 2.6(1) K after field cooling for an external magnetic 
field $\mu_0H=1$ T (a), 2 T (b), 3 T (c), and 4 T (d) (raw data). 
The crystal $c$-axis is aligned to neutron beam, 
and one of the \{1,0,0\} in-plane axis is aligned to the horizontal
axis in the plane perpendicular to the field. 
\label{sans}}
\end{figure}

Since we can reasonably assume that implanted muons (with 4 MeV) stop randomly on the 
length scale of FLL (which is 20--200 nm over the relevant field range), 
TF-\msr\ signal provides a random sampling of the internal field distribution $B({\bf r})$,
$\hat{P}(t) =\exp(-\sigma_p^2t^2)\int_{-\infty}^\infty 
n(B)\exp(i\gamma_\mu Bt-i\phi)dB,\label{Pt}$, 
with $n(B) = \langle\delta(B-B({\bf r}))\rangle_{\bf r}$,
where $\sigma_p$ is the additional relaxation due to random flux pinning,
$n(B)$ is the spectral density for the internal field defined as a spatial
average ($\langle\:\rangle_{\bf r}$) of the delta function over a unit cell of FLL,  
$\gamma_\mu$ is the muon gyromagnetic ratio (= 2$\pi\times 135.53$ MHz/T), and
$\phi$ is the initial phase of rotation.\cite{Brandt:88}
These equations indicate that the real amplitude of the Fourier transformed muon spin 
precession signal corresponds to the spectral density, $n(B)$. 
In the modified London model, which is a good approximation to evaluate $B({\bf r})$
at relatively lower external fields ($\mu_0H\ll H_{c2}$)\cite{Fesenko:93}, $B({\bf r})$ is 
given by a sum of the magnetic induction from isolated vortices to yield
\begin{eqnarray}
B({\bf r}) &=& \sum_{\bf K}b({\bf K})\exp(-i{\bf K}\cdot{\bf r})\label{mLondon}\\
b({\bf K}) &=& B_0\frac{\exp(-K^2\xi_v^2)}
{1+K^2\lambda^2+(\overline{n}_{xxyy}K^4+\overline{d}K_x^2K_y^2)\lambda^4}\:,\label{SpFou}
\end{eqnarray}
where ${\bf K}$ are the vortex reciprocal lattice vectors,
$B_0$ ($\simeq \mu_0H$) is the average internal field, $\lambda$ is the {\it effective}
London penetration depth, \xv\ is the cutoff parameter for the magnetic field
distribution near the vortex cores, and $\overline{n}_{xxyy}$ and $\overline{d}$ are dimensionless 
parameters arising from the nonlocal corrections with
the term proportional to $K_x^2K_y^2$ controlling the fourfold anisotropy:
the overline denotes an averaging of the parameter over the polar angle of
the symmetry axis, as it is anticipated for a mosaic of crystals.\cite{Kogan:97a}  
It is evident in Eq.(\ref{mLondon}) that $b({\bf K})$ is a spatial Fourier component of $B({\bf r})$.
The model predicts an asymmetric field profile for $n(B)$ characterized by a negatively 
shifted sharp cusp due to the van Hove singularity associated with the saddle points of $B({\bf r})$
and an asymptotic tail towards higher fields where the maximal field is determined
by $B(|{\bf r}-{\bf r}_i|\sim\xi_v)$ (with ${\bf r}_i$ being the vortex centers). 

Examples of the fast Fourier transform (FFT) of \msr\ spectra 
at several external magnetic
fields are shown in Fig.~\ref{fftall}, where one can readily observe the asymmetric
lineshape with slight smearing due to the random pinning of vortices, 
random local fields from nuclear moments, and the limited time
window for FFT ($\sim4$ $\mu$s: note that such smearing is 
irrelevant for the analysis in time domain).
Based on the least-square method with appropriate consideration
of the statistical uncertainty, the \msr\ spectra are compared in time domain with
those calculated by using Eqs.(\ref{mLondon})--(\ref{SpFou}) to deduce a set of parameters,
$\sigma_p$, $\lambda$, $\xi_v$, $\overline{n}_{xxyy}$, and $\overline{d}$: a 
typical example is shown in Fig.~\ref{fftall}. 
For the reconstruction of $B({\bf r})$ from $n(B)$, one needs to 
know the two-dimensional structure of FLL obtained from other 
experimental techniques including SANS.
In particular, it is noticeable in Fig.~\ref{fftall} that
the peak due to the saddle points exhibits a steep change with increasing
external field from 2 T to 3 T, suggesting an anomaly in the FLL structure.
As shown below, this speculation is confirmed by SANS measurements over 
the region of relevant magnetic field. 

\begin{figure}[h]
\includegraphics[width=1.0\linewidth]{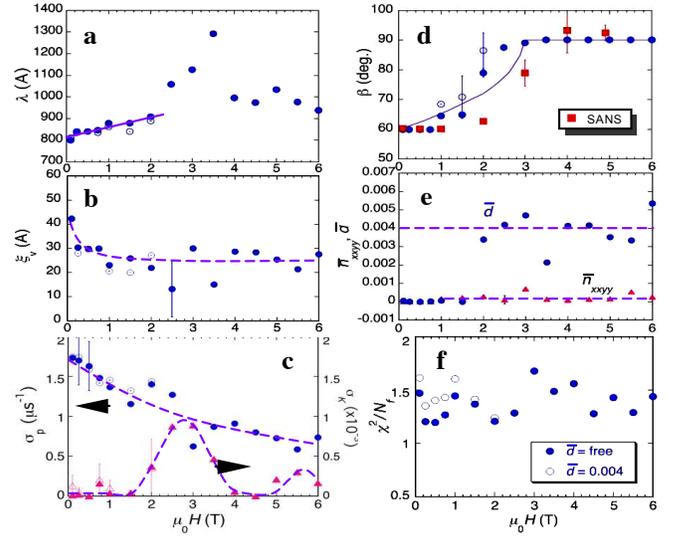}%
\caption{Magnetic field dependence of various parameters to describe the
field distribution $B({\bf r})$ obtained by analyzing TF-\msr\ spectra
using the modified London model: (a) the effective London 
penetration depth ($\lambda$), (b) the cutoff parameter (\xv), (c) the enhanced
relaxation rate due to local disorder of FLL ($\sigma_p$), and 
the mean width for the long-range distortion ($\sigma_{K}$), (d) the apex angle
of the rhombic unit cell in the real space (together with the SANS data), 
and (e) the dimensionless parameters
to describe anisotropy of supercurrent density around the flux lines ($\overline{n}_{xxyy}$, 
$\overline{d}$).
Error-bars in (a)--(e) show statistical errors (those not shown are smaller than the
symbol size), whereas the scatter of points, particularly for 2--4 T, may represent
systematic uncertainty (see text).  
The reduced chi-squares are shown in (f).  The open symbols in (a)--(d) and
(f) (only for $\mu_0H\le 2$ T) are those obtained by fixing $\overline{d}$ to 0.004. 
Dashed lines are merely intended to guide for eyes.  [See text for the solid lines in
(a) and (d).]
\label{msr}}
\end{figure}

In a SANS experiment, a FLL gives rise to
Bragg reflections at reciprocal lattice points, ${\bf K}=l{\bf u}+m{\bf v}={\bf K}_{lm}$. 
The intensity
of a single reflection $(l,m)$ is given by $|b({\bf K})|^2/|{\bf K}_{lm}|$. Considering
Eq.(\ref{SpFou}), it is reduced to $I_{10}\propto d_{10}\Phi_0^2/\lambda^4$
for a first order reflection, where $K_{10}=2\pi/d_{10}$ with $d_{10}$
being the FLL interval ($=\sqrt{3}a_0/2$ for a hexagonal FLL) and $\Phi_0$ is the flux quantum.
Fig.~\ref{sans} shows the typical SANS-J data at 2.6(1) K after subtracting background
data obtained at 30 K ($>T_c$, which is far below $T_M$), where 
scattered neutrons from an incident
beam with a mean wavelength $\lambda_0=6.5$ \AA\ 
($\Delta\lambda_0/\lambda_0\simeq$0.13 FWHM, collimated within a diameter of
2 mm) are detected by a position-sensitive detector (consisting of 128$\times$128
pixels with each pixel size being $4.95\times5.08$ mm$^2$, covering a sensitive
area of 58 cm ) at a distance of 10 m (1--2 T) or 4 m (3--4 T) from the specimen. 
One of the crystal $c$-axes and external magnetic field are aligned together 
to the neutron beam.
The cryostat (including the magnet) and specimen are rotated slightly 
($\pm0.2^\circ$, in several steps) around the vertical
axis to obtain a sum of diffraction patterns over Bragg angles for 
both triangular and square FLL structure.
It is clear in Fig.~\ref{sans} that there is a difference in the
pattern of angles between data at 1--2 T and those at 3--4 T.
Since the patterns at lower fields suggests an overlap of two 
(possibly distorted) hexagonal domains due to the tetragonal crystal structure,
their change into the nearly four-fold symmetric pattern at higher fields
demonstrate the occurrence of triangular-to-square lattice
transformation at 2--3 T. 
The field range of transformation is considerably 
higher than that in V$_3$Si, where the transition to square lattice is nearly complete at 
1.5 T (with $H\parallel$\{0,0,1\}).\cite{Yethiraj:99}
As shown in Fig.~\ref{msr}d, the apex angle of the real space unit cell is evaluated to be indeed 
close to 60$^\circ$ and 90$^\circ$ for the respective data.
Furthermore, the spots on the \{1,0,0\} axis
at high fields (Figs.~\ref{sans}c, d) is absent
at lower fields. This clearly indicates that the diagonal of the real space
rhombic unit cell, which is aligned to \{1,1,0\} or \{1,$\overline{1}$,0\} direction
in each domain, rotates to a \{1,0,0\} direction with
increasing field. A similar domain structure at lower field has been observed
in V$_3$Si.\cite{Yethiraj:99,Christen:85}

\begin{figure}[h]
\includegraphics[width=1.0\linewidth]{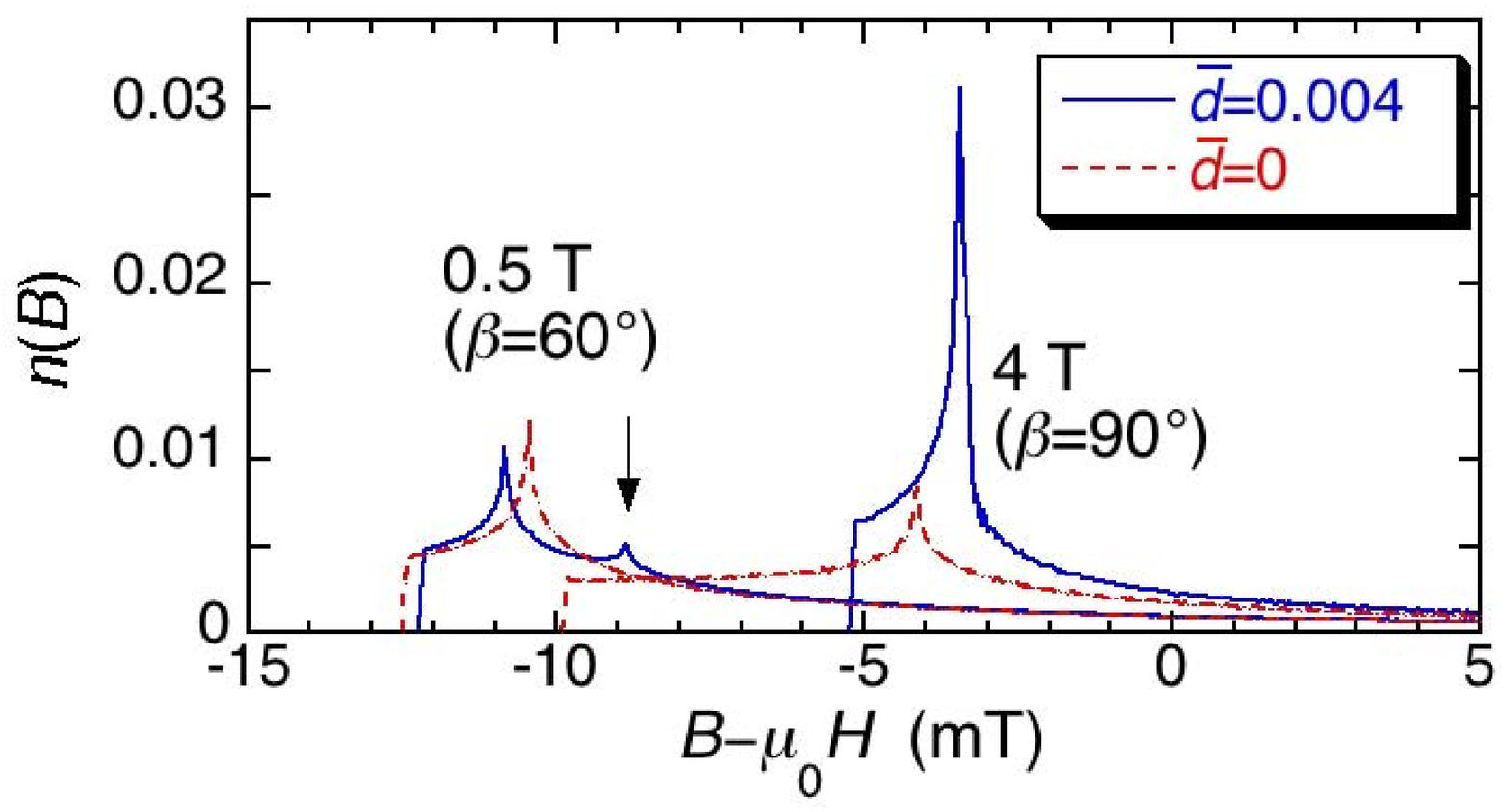}%
\caption{Spectral density distribution $n(B)$ calculated for a hexagonal FLL
with ($\overline{d}=0.004$) or without ($\overline{d}=0$) non-local correction 
under an external field of 0.5 T (where $\lambda=83.5$ nm, $\xi_v=3.1$ nm,
and $\overline{n}_{xxyy}=0$). The sharp peaks are due to van Hove singularities arising from 
turning points in $B({\bf r})$, where $\vec\nabla B({\bf
r}) = {\bf 0}$.  An additional peak (labeled by arrow) appears for the former case
because of the incommensurability between triangular FLL and squared shape 
of $B({\bf r})$ around a vortex.
Those for a square FLL at 4 T are shown
for comparison ($\lambda=100.0$ nm, $\xi_v=2.8$ nm, and $\overline{n}_{xxyy}=0$).
\label{nbd004}}
\end{figure}

Once the FLL structure is resolved by SANS measurements, high quality \msr\ spectra
can provide rich information on the mixed state though the detailed mapping
of $B({\bf r})$; as it has been demonstrated in the earlier works\cite{Sonier:04b,Ohishi:02},
the lineshape strongly depends on $\lambda$, $\xi_v$, $\overline{n}_{xxyy}$ and $\overline{d}$ 
in Eq.(\ref{SpFou}) and thereby they can be deduced by fitting analysis without
much ambiguity.  In particular, as shown in  Fig.~\ref{nbd004}, the spectral weight for the 
lower field side of the van Hove singularity strongly depends on the magnitude of 
$\overline{d}$ for the square FLL.  Besides these, 
we need to introduce two more parameters to describe the reciprocal vectors which undergo 
transformation: the apex angle, $\beta$,
and the mean width for the FLL distortion, $\sigma_{K}$,\cite{Kadono:01}
to describe the fluctuation of the reciprocal vectors,
${\bf K}=l(1+\delta_l){\bf u}+m(1+\delta_m){\bf v}$ 
with a probability distribution function,
$P(\delta_k)\propto\exp(-\delta_k^2/\sigma_{K}^2)$ $(k=l,m)$.
The last parameter effectively serves to add a Lorentzian-type broadening
to $n(B)$ and thereby discernible from the Gaussian broadening due to 
local disorder (which is described by $\sigma_p$). The randomness associated
with the mosaic crystals is presumed to be represented by $\sigma_K$. 
Fig.~\ref{msr} shows the field dependence of those parameters, 
where (a) $\lambda$, (b) $\xi_v$, (e) $\overline{n}_{xxyy}$ and $\overline{d}$ are directly related with
the local structure of supercurrent around a single flux line, whereas others
are with the morphology of the FLL. It is inferred from the
behavior of $\sigma_{K}$ (Fig.~\ref{msr}c)
that a strong long-range distortion occurs over a field range of 2--4 T,
where the FLL structure undergoes transformation between triangular
and square lattice. The anomalous enhancement of $\lambda$ observed in 
Fig.~\ref{msr}a is probably due to the artifact associated with this 
distortion; the reduced chi-square exceeds 1.5 around 3 T while it is
typically 1.2--1.4 for $\mu_0H<2$ T or $\mu_0H>$4 T (see Fig.~\ref{msr}f).
Similar scattering of the parameter value is also noticeable
in $\xi_v$, $\sigma_p$, and $\overline{d}$ over the relevant field range,
which may represent the systematic uncertainty due to the distortion. 
However, except for these anomalies, we can observe a clear trend in
all of the parameters. Namely, $\lambda$ exhibits a gradual increase with
increasing magnetic field below $\sim2$ T (which we discuss below in more detail), 
while $\xi_v$ tends to decrease slightly at
lower fields and levels off above $\sim2$ T. The behavior of
$\xi_v$, which to some extent represents that of the effective vortex core radius,
is commonly found in various superconductors, which is attributed
to the stronger overlapping of supercurrents around vortex center
and intervortex transfer of quasiparticles.\cite{Sonier:04}
The magnitude of \xv\ at lower field is in good agreement with the
Ginzburg-Landau coherence length, $\xi_{\rm GL}\simeq39$ \AA\
deduced from the relation $ H_{c2}=\Phi_0/2\pi\xi_{\rm GL}^2$
[with $H_{c2}$(2 K) being assumed to be 22 T].
As is anticipated from the SANS result, $\beta$ 
exhibits a steep change from 60$^\circ$ to 90$^\circ$ around 2--3 T. 
The behavior is qualitatively in good 
agreement with the prediction of the non-local model 
(solid curve\cite{Kogan:97a,Kogan:97b}, where the critical field is assumed to be 3 T). 
Thus, it indicates that the 
anisotropic flow of the supercurrent due to the nonlocal effect 
is the primary origin of the square-lattice formation at higher fields.
The magnitude of $\overline{d}$ ($\simeq4\times10^{-3}$), which should be regarded as a
lower boundary considering the fact that it is reduced by
an averaging over a varying symmetry axis,\cite{Kogan:97a} is considerably larger
than that observed in V$_3$Si ($\simeq0.3$--$1\times10^{-3}$),\cite{Sonier:04b}
suggesting a stronger anisotropy in the Fermi surface. 
The spin relaxation due to the
random local distortion ($\sigma_p$) exhibits a gradual decrease with 
increasing field: this is qualitatively understood as a result of stronger
overlap of flux lines at higher field that 
makes less room for flux lines to deviate
away from the regular position. Unfortunately, the SANS data are available
only up to 5 T due to experimental conditions, and therefore the origin 
of the weak anomaly around 5.5 T (see Fig.~\ref{msr}c) is not identified at this stage. 

It is Interesting to note that the anisotropy parameter $\overline{d}$ is sharply reduced to 
zero below $\sim$2 T: a similar behavior is  
reported for V$_3$Si.\cite{Sonier:04b}   As shown in Fig.~\ref{msr}f,
a comparative analysis with $\overline{d}$ fixed to 0.004 ($\sim$ an average for $\mu_0H> 2$ T) 
yields systematic increase of chi-squares, while little variation is found for other parameters.
This is attributed to an additional cusp in $n(B)$ (see Fig.~\ref{nbd004}) corresponding to the
saddle points relatively near the vortex cores where $B({\bf r})$ is influenced
by the squared shape of supercurrent flow.  Our result indicates that 
the field profile over the field range of hexagonal FLL may be better 
described by the {\sl isotropic} London model. 

Now, we discuss the origin of gradual increase of $\lambda$ in Fig.~\ref{msr}a.
Our systematic study on the magnetic field dependence of $\lambda$
in various superconductors indicates that the gradient ($d\lambda/dH$) 
provides information on the low energy quasiparticle excitations.\cite{Kadono:04}
In the mixed state, the quasiparticle momentum ${\bf v}_{\rm F}$ is shifted by the flow of
supercurrent ${\bf v}_{\rm s}$ around the vortices due to a semi-classical Doppler shift,
leading to a shift of the quasiparticle energy spectrum to an amount 
$\varepsilon=m{\bf v}_{\rm F}\cdot {\bf v}_{\rm s}$. When the 
superconducting order parameter has a nodal structure or anisotropy
(including multi-gap), the Cooper
pairs with a gap energy of less than $\varepsilon$
can be broken by the Doppler shift,\cite{Volovik:93,Nakai:04} leading to
an enhancement of $\lambda$.  Using a normalized magnetic field, $h=H/H_{c2}$, 
we can express the field dependence of $\lambda$,
\begin{equation}
\lambda(h)=\lambda(0)[1+\eta\cdot h],\label{eta}
\end{equation}
where $\eta$ is the dimensionless parameter to represent the gradient
(the magnitude of pair breaking effect).
Since $\lambda^{-2}(h)=4\pi e^2n_{\rm s}(h)/m^*c^2$ 
that is determined by the superfluid density, $n_s(h)$, 
the reduction of $n_s(h)$ due to the Doppler shift
would lead to the increase of $\lambda(h)$.
A fitting analysis of the data in Fig.~\ref{msr}a by Eq.(\ref{eta})
below 2 T yields $\eta=1.24(2)$ with $\lambda(0)=818(1)$ \AA, where the result 
is shown by a solid line in Fig.~\ref{msr}a.
Considering that no clear evidence is reported for the presence of nodes in the
order parameter, the situation in \nbsn\ is close to  
the case of MgB$_2$ where a similar trend of $\lambda(h)$ with $\eta\simeq1.3$ 
was observed.\cite{Ohishi:03} Since \msr\ measurements are made at finite 
temperature ($T$), the Cooper pairs having the energy near the gap minimum
(or a smaller gap) $\Delta_s$ would be susceptible to excitation
due to the Doppler shift when $\Delta_s\le k_BT$.\cite{Nakai:04} 
Interestingly, an earlier report on
electronic Raman scattering experiment suggests a relatively strong gap anisotropy
($\sim 20$\%) in Nb$_3$Sn, whereas no such anomaly seems to exist 
for V$_3$Si.\cite{Dierker:83}
Meanwhile, a very recent specific heat measurement in \nbsn\ demonstrated the presence 
of residual excitation below $\sim0.3T_c$, which they attributed to a double-gapped 
order parameter with $\Delta_s=0.61(5)$ meV.\cite{Guritanu:04} 
Our result is another, yet microscopic evidence for the anisotropic 
(or multi-gapped) $s$-wave pairing for the order parameter of \nbsn. The anisotropy 
of the Fermi surface revealed by FLL transformation may be related with the
structure of the order parameter. Further study using the techniques sensitive
to gap anisotropy would be helpful to determine the gap structure in full detail.

In summary, we have obtained detailed mapping of the local magnetic field
in the mixed state of \nbsn\ using \msr\ and SANS, 
from which we extracted fundamental length scales of
superconductivity. Unlike V$_3$Si, the field-induced enhancement of the effective
penetration depth strongly suggests that the order parameter in \nbsn\ has
a structure, either anisotropic or multi-gapped.

We would like to thank the staff of TRIUMF for their technical
support during the \msr\ experiment.
This work was partially supported by a
Grant-in-Aid for Scientific Research on Priority Areas and
a Grant-in-Aid for Creative Scientific Research from the Ministry of
Education, Culture, Sports, Science and Technology of Japan.

\end{document}